\documentclass[10pt]{iopart}
\usepackage{graphicx}
\usepackage{hyperref}

\begin{document}

\title{Scalable chip-based 3D ion traps}

\author{Elena Jordan$^{1,2}$, Malte Brinkmann$^1$, Alexandre Didier$^1$, Erik Jansson$^1$, Martin Steinel$^1$, Nils Huntemann$^1$, Hu Shao$^1$, Hendrik Siebeneich$^3$, Christof Wunderlich$^3$, Michael Johanning$^{3,4}$, Tanja E. Mehlst\"aubler$^{1,5,6}$}

\address{$^1$ Physikalisch-Technische Bundesanstalt, Bundesallee 100, 38116 Braunschweig, Germany\\
				$^2$ Deutsches Zentrum f\"ur Luft- und Raumfahrt (DLR) Institut f\"ur Satellitengeod\"asie und Inertialsensorik, Callinstra\ss e 30b, 30167 Hannover, Germany\\ 
				$^3$ Naturwissenschaftlich-Technische Fakult\"at, Universit\"at Siegen, Walter-Flex-Straße 3, 57068 Siegen, Germany\\
				$^4$ eleQtron GmbH, Heeserstr. 5, 57072 Siegen, Germany\\
				$^5$ Institut f\"ur Quantenoptik, Leibniz Universit\"at Hannover, Welfengarten 1, 30167 Hannover, Germany\\
				$^6$ Laboratory of Nano and Quantum Engineering, Leibniz Universit\"at Hannover, Schneiderberg 39, 30167 Hannover, Germany} 

\ead{elena.jordan@ptb.de}
\vspace{10pt}

\begin{abstract}
Ion traps are used for a wide range of applications from metrology to quantum simulations and quantum information processing. Microfabricated chip-based 3D ion traps are scalable to store many ions for the realization of a large number of qubits, provide deep trapping potentials compared to surface traps, and very good shielding from external electric fields. In this work, we give an overview of our recent developments on chip-based 3D ion traps. Different types of chip materials, the integration of electronic filter components on-chip and compact electrical connections in vacuum are discussed. Further, based on finite element method (FEM) simulations, we discuss how integrating micro-optics in 3D ion traps is possible without disturbing the trapped ions.    
\end{abstract}

%
\vspace{2pc}
\noindent{\it ion traps, integrated optics}
%
%
%
\ioptwocol

\section{Introduction}
		
Ion traps are the heart of physics packages for applications in quantum metrology, quantum computation and simulation \cite{Ludl2015, Bruz2019, Dege2017}.  In particular, scalable ion traps are indispensable for ion-based quantum computing processors and scalable optical clocks \cite{Hild2022, Debn2016, Ryan2022, Ring2022, Bluem2021, Brew2019, Hunt2016}. At PTB, we have developed scalable chip-based 3D ion traps with high manufacturing precision which were benchmarked for ion precision spectroscopy with low systematic errors \cite{Kell2019, Haus2024}. Here, we report on developments in our production line of chip-based 3D ion traps, ranging from printed circuit board (PCB) assemblies to high-precision traps that can be manufactured out of sapphire, diamond, or ceramics. Within the \textit{opticlock} project \cite{Ritt2020}, we have developed a compact ion trap for multi-ion clock operation. In the \textit{IDEAL} project \cite{ideal}, the design has been further developed into a diamond-based two chip version with integrated micro-optics for simultaneous addressing and readout of ions in multiple trap segments.  
 
Integrated optics, such as fibers, lenses, mirrors, waveguides and photodiodes, have previously been investigated for scalable or enhanced light delivery and state detection \cite{Fern2023, Young2014, Day2021, Meht2020, Niff2020,  Aran2020, Setz2021, Reen2022}. They have also proven to be useful for the integration of an optical cavity with an ion trap for scalable quantum networks \cite{Taka2013, Taka2017, Lee2019, Fern2023, Tell2023}. Here, we discuss how integrated micro-optics influence the trapping fields by means of FEM calculations and show that the chip design allows for the installation of micro-optics within the trap assembly with little distortion of the electric trapping fields. 

In Section 2, we give an overview of different types of 3D chip traps produced in our group. In Section 3, we introduce compact electrical connections to segmented traps, filter electronics, and estimate how Johnson noise can contribute to the heating rate of the ions. In Section 4, we use FEM simulations to examine the trapping fields of 3D chip ion traps in the presence of integrated optics. 

\begin{figure}
	\centering
		\includegraphics[scale = 0.32]{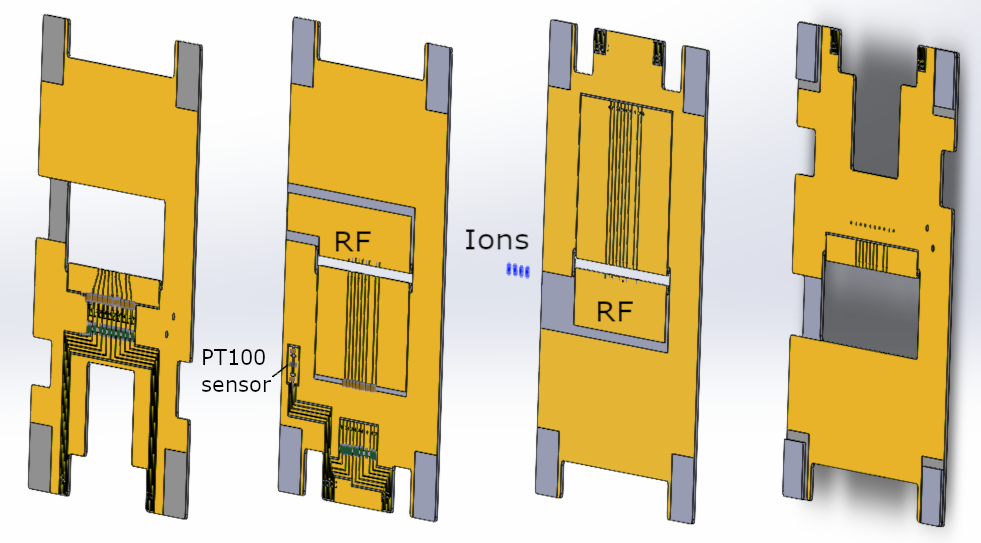}      
		\caption{\label{Fig1}Trap design with 4 chips, the two central ones provide the RF quadrupole field and the axial trapping potential, the two outer chips the compensation fields.}
\end{figure}

\begin{figure}
	\centering
		\includegraphics[scale = 0.16]{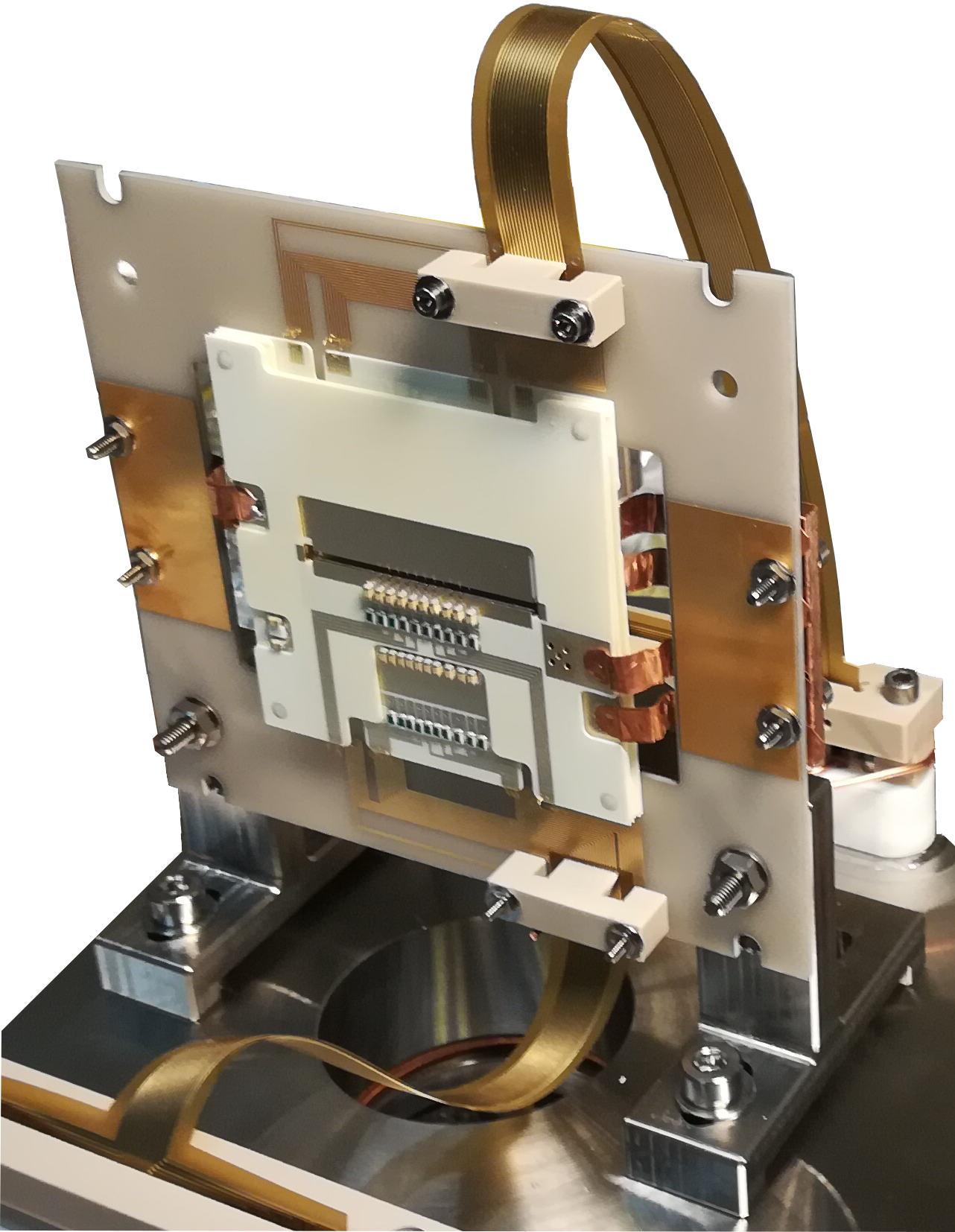}      
		\caption{\label{Fig2}Chip ion trap made from PCB material. The RC filter electronics and temperature sensors are positioned on the trap chips, the RF potentials are supplied via copper strips. The trap is connected to an aluminium nitride (AlN) carrier board which is used to mount the trap assembly on a CF flange. For the DC voltages a flexible ribbon with printed leads is clamped onto the carrier board.}
\end{figure}

\section{Types of four chip ion traps and materials}
For precision spectroscopy and optical clocks, we have developed and optimized a 3D chip ion trap \cite{Burg2019, Pyka2014, Hers2012}. The ion trap has been operated for precision time and frequency metrology \cite{Kell2019, Stei2023, Haus2024}, spectroscopy \cite{Drei2022, Fuer2020}, and simulations of many-body physics \cite{Kiet2021, Kiet2017}.  It consists of four chips, two radio frequency (RF) chips for the radial and axial trapping of the ions, and two chips for the compensation of stray electric fields as shown in figure \ref{Fig1}. The chips have a 1 mm wide through slot in the center and are assembled to a stack, where the precise alignment of the chips is crucial for low micromotion and low heating rates in the trap \cite{Kalin2021, Hers2012}. The ions are trapped in the center of the stack between the RF chips that have a distance of 1 mm. This results in an ion to electrode distance of 0.7 mm. RC low pass filters are placed directly on the trap chips to minimize the pickup of electrical noise on the direct current (DC) electrodes. The filter electronics can be chosen depending on the application. For the ion traps discussed here, we chose a cutoff frequency between 100 Hz and 1.1 kHz to suppress noise at the secular frequencies of the ions. PT100 temperature sensors are placed on the RF chips for an in situ temperature measurement of the trap. 

\subsection*{Ion traps made from a PCB material} 

Chip ion traps are not only useful for precision metrology, but can be adapted for different applications, e.g. the optimal number and size of segments and electrode geometries vary depending on the demands. To test new designs affordable prototype traps are needed that are available with short turnaround times. Ion traps made from a PCB material which is optimized for RF applications\footnote{such as Rogers 4000} are ideal for this purpose \cite{Brown2007, Ceti2007, Kumph2016, Pyka2014}. They are available on the same time scale as any PCB and the cost is low compared to ceramic or crystalline chips. Furthermore, PCB chips are robust and easy to handle. At the same time, they offer very good RF properties with a loss tangent of $\tan(\delta) = 31\times 10^{-4}$ at 2.5 GHz at $23\ ^{\circ}$C and can store ions with low micromotion \cite{Pyka2014}. Heating rates as low as $\dot{\bar{n}}=1.2 \pm 0.4$ s$^{-1}$ and $\dot{\bar{n}}=1.3 \pm 0.5$ s$^{-1}$ for radial modes at $\nu_{\mathrm{sec,1}}= 440$ kHz and $\nu_{\mathrm{sec,2}}=492$ kHz, respectively, have been measured in some of these traps \cite{KellerProceedings2015}. An example of an ion trap made from PCB material is shown in figure \ref{Fig2}.

\begin{figure}
	\centering
		\includegraphics[scale= 0.153]{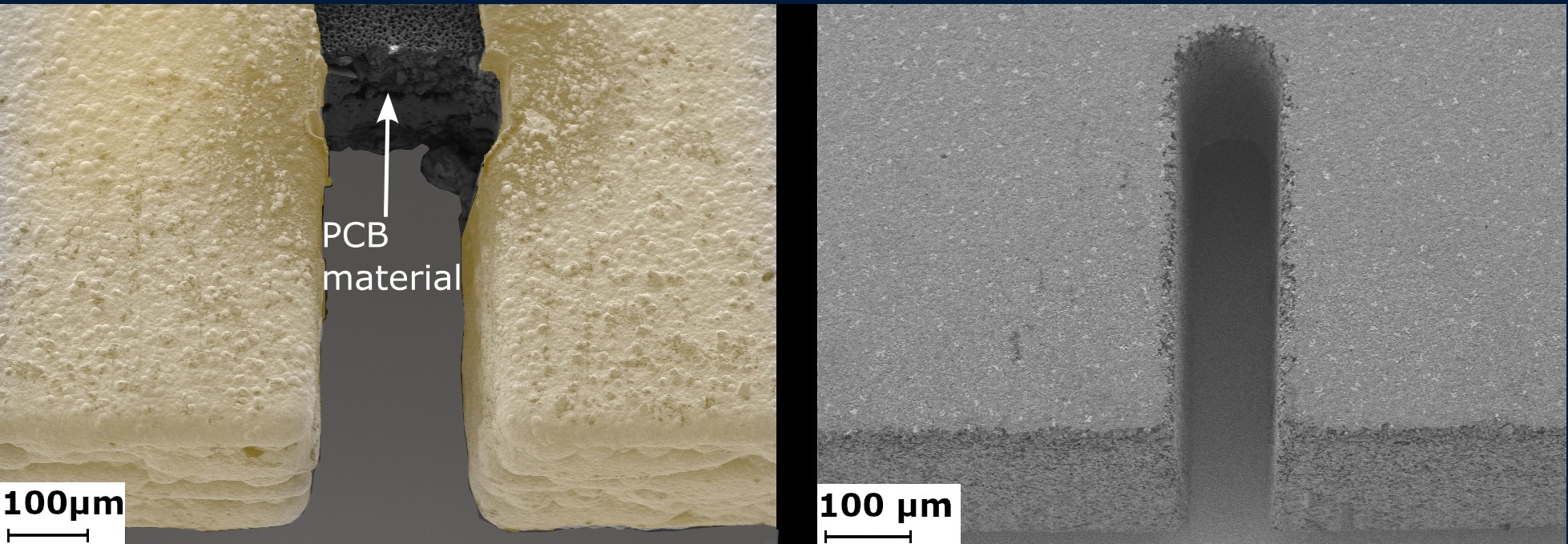}    
		\caption{\label{fig:sem}left: SEM image of a PCB trap chip. The image shows part of two trap segments with gold coating (colored yellow) that are isolated by a slit and the bare PCB board material which is visible in the center of the image.  right: SEM image of an uncoated AlN trap chip. The image shows part of two trap segments, separated by a slit.}
\end{figure}

\subsection*{RF traps made from aluminium nitride ceramics} 

RF traps made from aluminium nitride (AlN) ceramics have excellent RF properties with a loss tangent of $\tan(\delta) = 3\times 10^{-4}$ at 1 MHz at $20\ ^{\circ}$C and about a factor 250 higher thermal conductivity compared to PCB traps \cite{Nord2020}. Another advantage is the stiffness of the ceramic chips which makes narrower manufacturing tolerances possible but requires advanced fabrication processes. A comparison of scanning electron microscope (SEM) images in figure \ref{fig:sem} demonstrates that the AlN ceramic chip has smoother surfaces and sharper, more precise edges. The manufacturing tolerances are critical for the performance of the traps, in particular, if low micromotion and well-controlled heat management is desired \cite{Kell2019,  Nord2020, KellerProceedings2015}. An example of a trap made from AlN ceramics with filter electronics placed on the trap chips is shown in figure \ref{Fig43}. The chips are monolithically laser cut from AlN wafers, coated with titanium as an adhesion layer and gold for the electrode surface. The RC filters (detail in figure \ref{Fig43}) reduce electronic noise on the electrodes for the 10 trap segments.

Heating rates of the ions in this trap have been measured in two different experimental setups. The observed heating rate in one setup is $\dot{\bar{n}} = 0.56\pm 0.06 $ s$^{-1}$ per ion in a four ion crystal for the radial center of mass mode at frequencies $\nu_{\mathrm{sec}}$ between  615 kHz and 635 kHz \cite{Kalin2021}. The heating rates measured with a single ion in another setup at three different secular frequencies are $\dot{\bar{n}} = 0.76\pm0.24$ s$^{-1}$, $\dot{\bar{n}} =0.95\pm0.24$ s$^{-1}$, and $\dot{\bar{n}} =0.53\pm0.15$ at a radial secular frequency $\nu_{\mathrm{sec}}$ of 597 kHz,  606 kHz, and 1,028 kHz, respectively \cite{Tabea2023}.

\subsection*{Compact AlN ceramic trap}

\begin{figure}
	\centering
		\includegraphics[scale = 0.15]{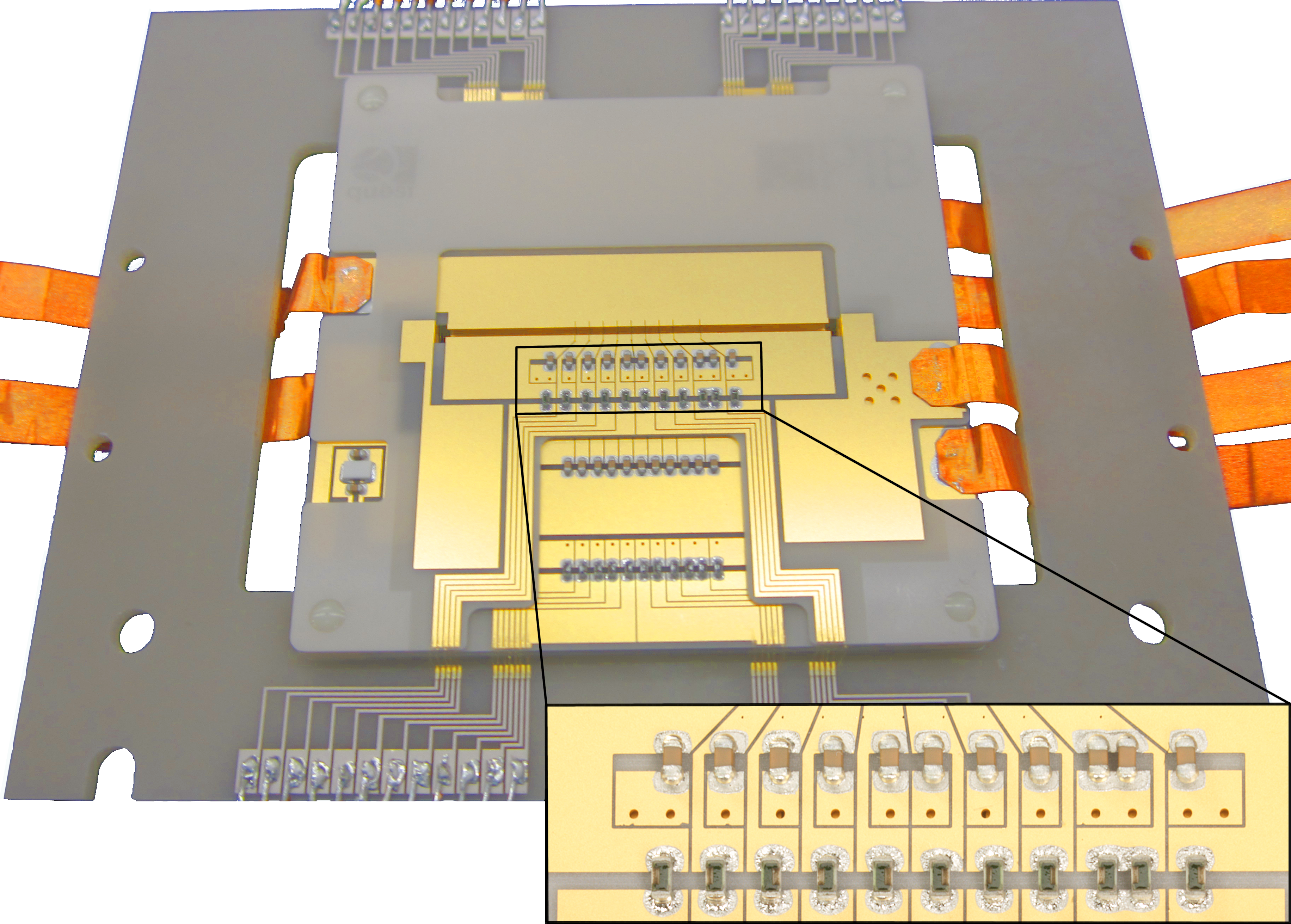}      
		\caption{\label{Fig43}Chip ion trap made from AlN ceramic chips. The RC filter electronics (detail) for the DC electrodes for axial trapping and for the compensation electrodes, and the temperature sensors are placed on the trap chips. The copper strips connect to the RF power supply. The trap consists of a stack of four chips.}
\end{figure}

\begin{figure}
\centering
		\includegraphics[scale =0.396]{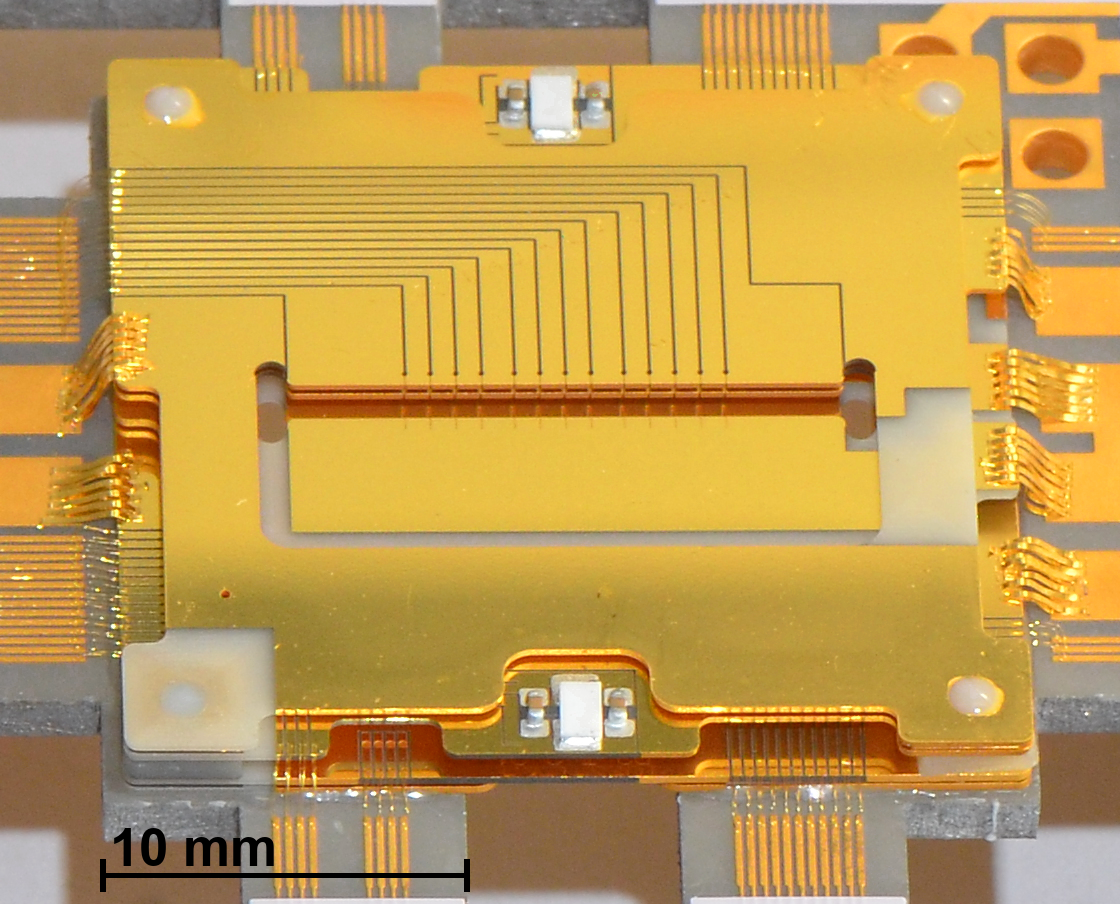}%
		\includegraphics[scale =0.450]{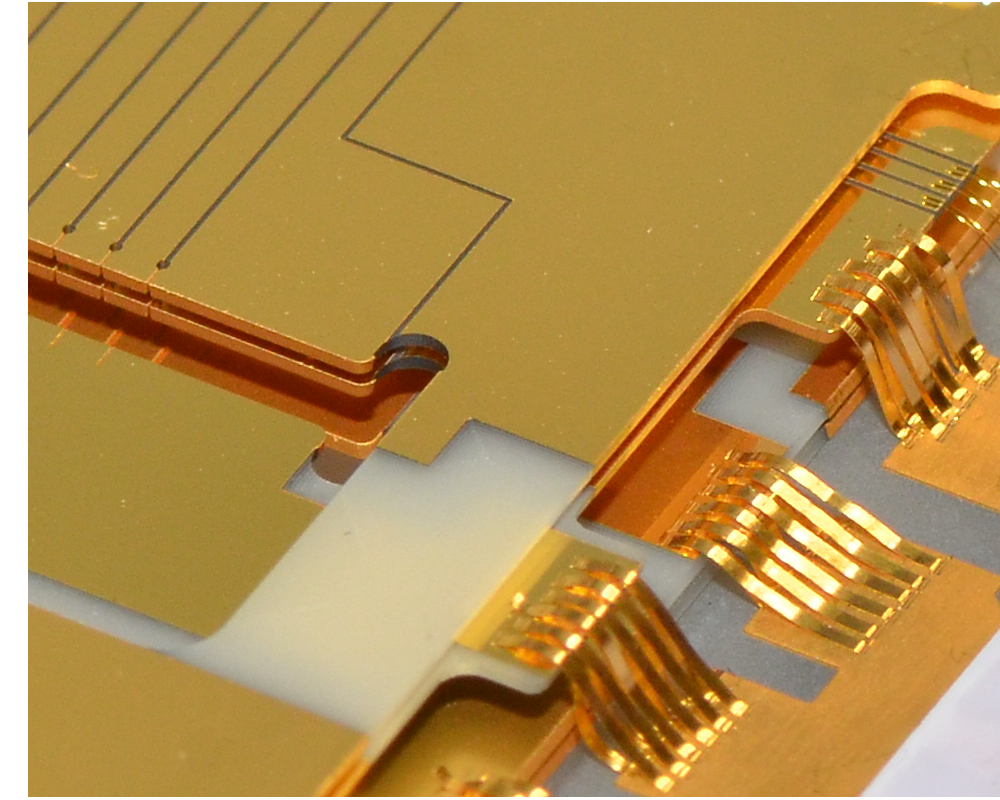}
		\includegraphics[scale =0.324] {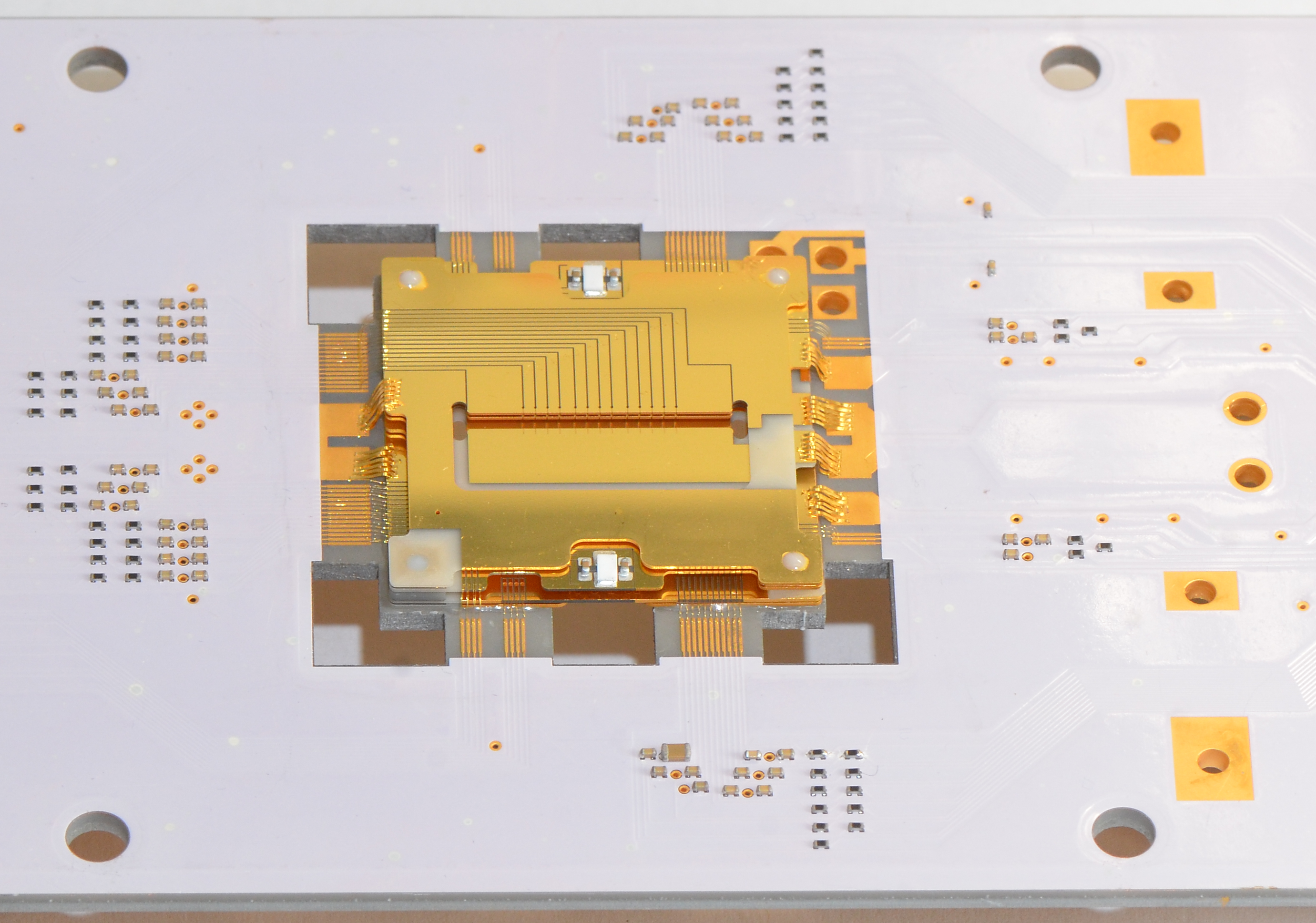}
		\caption{\label{Fig5}\textit{top left} Compact chip-based 3D ion trap with PT100 sensors on the RF chips for temperature measurements. \textit{top right} Detail showing the ribbon bonds for the RF connection. Each connection is made of twelve bonds with 250 $\mu$m width in two layers. \textit{bottom} Ion trap on a carrier board with filter electronics for the DC electrodes. }
\end{figure}

Within the \textit{opticlock} project \cite{Ritt2020} a more compact AlN ceramic trap (Fig. \ref{Fig5}) with 14 trap segments has been developed. The chips measure only 30 mm $\times$ 30 mm, instead of 50 mm $\times$ 50 mm  for the trap in figure \ref{Fig43}. The PT100 temperature sensors remain on the trap chips to deliver reliable precise in situ measurements of the temperature of the RF chips, while the RC low pass filters are placed on the carrier board of the trap. Ribbon bonds are used for the RF connections instead of soldered copper strips. The RF voltage was connected using 12 gold ribbon bonds with 250 $\mu$m width and 25 $\mu$m thickness (Fig. \ref{Fig5}) permitting 1500 V and 3 A current with a resistive heating comparable to the copper strips. 

\begin{table}
\small
		\begin{tabular}[h]{lcc}
		\hline
			Material			& Therm. conductivity &  Loss tangent\\
										&	(in W m$^{-1}$ K$^{-1}$)& ($\times 10^{-4}$)\\
		\hline
			Rogers 4350B	& 0.7									&	31 (@2.5 GHz) \\
			AlN ceramics		&	140 - 180										& 3	(@1 MHz)				\\
			Al$_2$O$_3$ ceramics	&	22 - 24					& 3 (@1 MHz)	 \\
			Fused silica	&	1.4										& $<$ 1 (@1 MHz)	\\
			Sapphire			&	40										& 0.3 (@10 GHz)		\\
			CVD Diamond		&	$>$ 1800							  & 10  (@1 MHz)	\\
			Si					& 145 										& 19 (@7 GHz )\\
			\hline
		\end{tabular}
	\caption{Comparison of materials for trap chips at room temperature and at 80 $^\circ$C for Rogers 4350B. References for the data for Rogers 4350B see Ref. \cite{Rogers}, for gray AlN ceramics see Ref. \cite{AlN}, for Al$_2$O$_3$ (96 \%) Ref. \cite{AlO}, for fused silica see Ref. \cite{fusedsilica}, for sapphire perpendicular to C axis Ref. \cite{sapphire}, for diamond see Ref. \cite{Ibar1997, Diamondmat}, and for Si see Ref. \cite{Maycock1967, Krupka2006}}.
	\label{tab:MaterialsForTrapChips}
\normalsize
\end{table}

\section{Connectorization, SMD filter electronics, and Johnson noise}
\subsection*{DC connections}
Scalability is important for applications that benefit from larger number of ions in the trap. For example, to gain higher signal to noise ratios in precision measurements, to study many-body physics, or to increase the number of qubits for quantum computing. Therefore, there is an increasing demand for scalable ion traps with well-controlled micromotion and multiple trap segments. For segmented ion traps, the cables supplying the DC voltages for the axial confinement need lots of room and can obstruct the laser access.

\begin{figure}
	\centering
		a)\includegraphics[scale = 0.18]{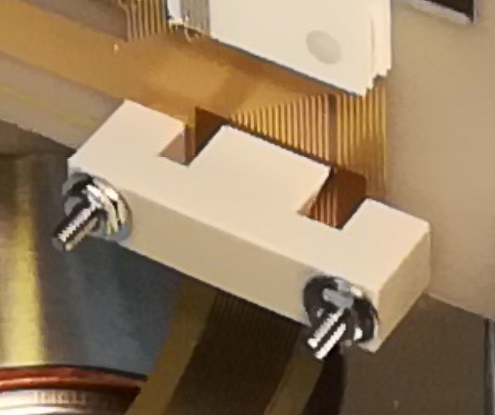}%
		b)\includegraphics[scale = 0.38]{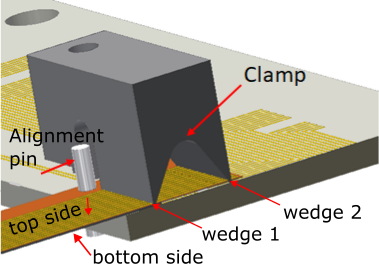}		
		\caption{\label{Fig3}a) Clamp connecting a flexible ribbon printed circuit board for the DC voltage supply on the carrier board (see Fig. \ref{Fig2} for complete trap setup). b)  Section view of the clamp. The alignment pins align the ribbon band to the pads on the board, the first wedge clamps down the conductor tracks on the bottom side, the second wedge clamps down the conductor tracks from the top side that are brought to the bottom via through-hole vias.}
\end{figure}

We have developed solder-free compact DC connections. Gold coated copper conductors with 200 $\mu$m width and 200 $\mu$m pitch on both sides of a flexible polyimide ribbon with 50 $\mu$m thickness are used. The outgassing behaviour of polyimide films is well studied and outgassing rates of the order of $1\times 10^{-9}\ \mathrm{Pa\  m}^3/(\mathrm{s\ m}^2)$ have been observed after baking \cite{Battes2017}. The flexible ribbon is aligned to the DC contact pads on the carrier board of the trap using two alignment pins and clamped down onto the carrier board as shown in figure \ref{Fig3}a. The tolerance for the position of the alignment pins is 50 $\mu$m. This ensures reliable and reproducible electrical connections. The conductors on the upper side of the ribbon are connected to the pads via through-hole vias through the band that are clamped onto a second row of contact pads. 

A section view of the clamp with the two wedges for clamping down the conductors on the top and bottom side of the ribbon onto the contact pads on the carrier board is shown in figure \ref{Fig3}b. The other end of the flexible ribbon is connected to a sub-D type vacuum feedthrough, see figure \ref{Fig2}. We used two ribbon cables with 37 conductor tracks. The design can be adjusted for different geometries of the contact pads and different numbers of conductors as needed for the respective trap. 

\subsection*{RC filter electronics}
RC low-pass filters are used to reduce the high frequency noise on the DC electrodes. We typically chose a resistance in the range 25--300 k$\Omega$ and a capacitance in the range 4.7 -- 38 nF. The SMD filter electronics are attached by reflow soldering. The flux contained in the solder paste is removed with a defluxer before the chips are cleaned for UHV applications \cite{LIGO}. The melting temperature of the solder (Kester EM907 Solder Paste\footnote{ https://www.kester.com/products/product/em907-solder-paste }) is $222\ ^\circ$C at atmospheric pressure according the data sheet. To see what the maximum bakeout temperature of the trap is, we perform tests with PCB trap chips with soldered SMD parts under reduced pressure at bakeout temperatures $120\ ^\circ$C, $160\ ^\circ$C, and $230\ ^\circ$C. 
 
The conclusion is that the traps withstand bakeouts at $160\ ^\circ$C. After baking the setup with the PCB trap for 4 weeks at $160\ ^\circ$C a pressure $<1\times 10^{-10}$ mbar was reached. The estimated pressure agreed well with the observed collision rate $\Gamma = (1.1 \pm 0.1)\times10^{-3}$ s$^{-1}$ ion$^{-1}$ indicated by reordering of the ion crystal \cite{Hank2019}. In the experiments using AlN ceramic traps (Fig. \ref{Fig43}) a collision rate between $\Gamma = (1.03 \pm 0.05)\times 10^{-3}$ s$^{-1}$ ion$^{-1}$ and $\Gamma =(2.9 \pm 0.3)\times 10^{-3}$ s$^{-1}$ ion$^{-1}$ is observed based on reordering of ion crystals with up to 16 ions.

\subsection*{Johnson Noise}
\begin{figure}
	\centering
		\includegraphics[scale = 0.8]{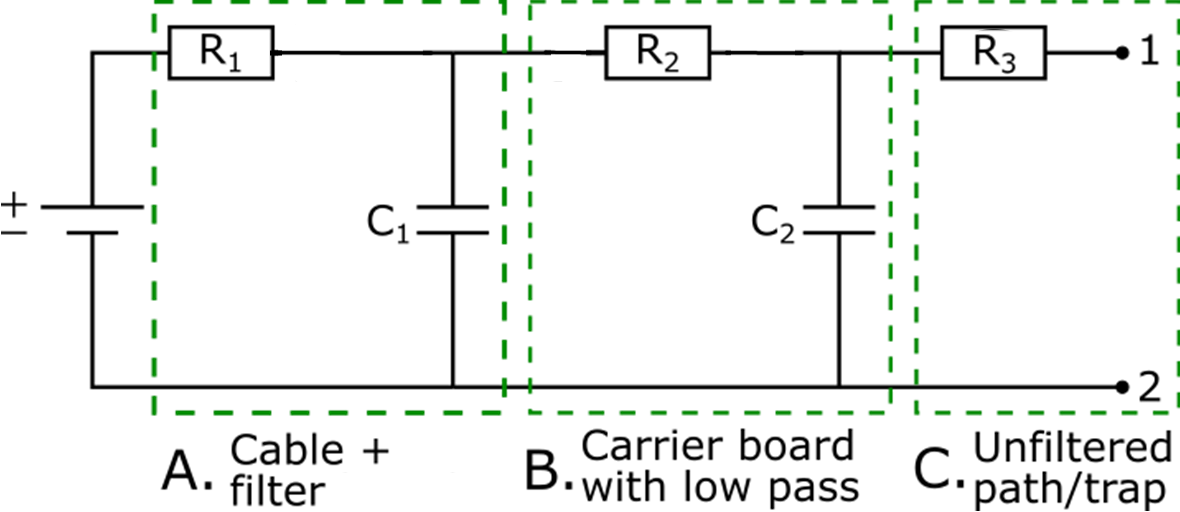}  
		\caption{\label{Fig:JohnsonNoise} Circuit diagram for the calculation of the Johnson noise on one DC electrode (1). Part A contains the resistance R$_1$ and capacitance C$_1$ of the cables and a first low pass filter. Part B contains the conductor tracks on the carrier board of the trap with low pass filters with resistance R$_2$ and capacitance C$_2$. Part C contains the conductor tracks on the trap chips that connect to the electrodes of the ion trap with R3. }
\end{figure}

Electric field noise on the trap electrodes in the frequency range of the secular frequency leads to heating of the ions. We estimate the contribution of the Johnson noise that arises from the thermal motion of charge carriers for room temperature operation, i.e. 300 K, for the example of one DC electrode of the compact AlN chip trap shown in figure \ref{Fig5}. The calculation is based on the reduced circuit diagram in figure \ref{Fig:JohnsonNoise} and takes into account the passive electrical elements on the circuit. Noise from the voltage source is not taken into account. Placing the RC filter electronics on the ion trap chips directly (Fig. \ref{Fig43}) limits the length of unfiltered sections of the conductor tracks to the DC electrodes.  When the filters are placed on the carrier board, the unfiltered section is prolonged. In the trap setup in figure \ref{Fig5} (bottom) the low pass filters for the DC electrodes are positioned on the carrier board. The calculation considers the resistance and capacitance of the trap electrode, conductor tracks on the trap chip and the carrier board, the SMD electronics for the RC low-pass filters, and the cables and filter between the voltage source and trap setup. 
The spectral density of the electric field noise due to Johnson noise depends on the effective real resistance $R$  between points 1 and 2 in the diagram (Fig. \ref{Fig:JohnsonNoise}) and is given by \cite{Brow2015}
\begin{equation}
S_V^J=4k_BTR(\omega, T),
\label{eq:JohnsonNoise}
\end{equation}
where $k_B$ is the Boltzmann constant, $T$ is the temperature, $R$ is the effective real resistance of the circuit, and $\omega$ the RF frequency. 
First, the effective resistance of the section before the trap electrode including the low pass filters is determined. Second, the effective resistance of the circuit after the filters. For the cables and filters between voltage source and experimental setup we estimate a resistance of $R_1 \approx 20~\Omega$ and capacitance $C_1 \approx 10$ $\mu$F. The low pass filters on the carrier board consist of resistors with $R_2 = 30$~k$\Omega$ and capacitors with $C_2 = 4.7$~nF resulting in a cutoff frequency of $\omega_{\mathrm{cutoff}} =2\pi \times 1.1$~kHz.  For the calculation of the effective resistance, we use $R_1 \ll R_2 $ and $C_1 \gg C_2$. With this, the effective resistance for a secular frequency of $\omega = 500$~kHz is \cite{Brow2015}

\begin{equation}
R_{\mathrm{eff}, 1} \approx \frac{R_2}{(\omega R_2 C_2)^2+1} \approx 153~\mathrm{m}\Omega.
\label{eq:Resistance}
\end{equation}

In addition, the conductors on the carrier board contribute with a resistance of about 33~m$\Omega$, a pair of two bond wires with a length of up to 3 mm contribute 67 m$\Omega$, and the trap electrodes themselves contribute a resistance of 418 m$\Omega$ to an effective resistance of the circuit after the low pass filters to the trap electrodes. In sum, we estimate the resistance of the longest path to be $R_{3} = 520$~m$\Omega$. 
The effective resistance between 1 and 2 in figure \ref{Fig:JohnsonNoise}, $R_{12}$, is the sum of the two contributions $R_{12}=R_{\mathrm{eff}, 1}+R_{3}= 673 $ m$\Omega$. For the spectral density of the electric field noise (Eq. \ref{eq:JohnsonNoise}) at room temperature for a secular frequency of $\nu_{\mathrm{sec}} = 500$~kHz we obtain 

\begin{equation}
 S_V = 1.1\times 10^{-20}\ \mathrm{V}^2 \ \mathrm{Hz}^{-1}.
\label{eq:noisedensity}
\end{equation}

From the noise spectral density, the heating rate for a single ${}^{171}$Yb$^+$ ion can be calculated with the relation \cite{Wine1998}
\begin{equation}
	\dot{n}=\frac{e^2}{4 m \hbar \nu_{\mathrm{sec}}}\frac{2S_V}{D^2}\approx 0.31	\ \mathrm{s}^{-1},
	\label{eq:heatingrate}
\end{equation}
where $e$ is the elementary charge, $\hbar$ is Planck's constant, $\nu_{\mathrm{sec}}$ is the secular frequency, $m$ the mass of the ion, and $D$ the characteristic distance of the trap electrodes to the ion. It is the separation that a parallel plate capacitor with a voltage $V$ applied would need to have, to result in the same electric field as, to lowest-order approximation, arises near trap center from the pair of trap electrodes with voltage $V$ applied \cite{Wine1987}. Here, it is determined from FEM simulations, resulting in $D=2.2$ mm. Due to screening effects, $D$ is considerably larger than the ion to electrode distance of 0.7 mm. The factor 2 in equation \ref{eq:heatingrate} accounts for the uncorrelated noise on the two trap electrodes that provide the DC potential. 
In conclusion, the estimated heating rate due to Johnson noise is up to 0.31 phonons per second at 500 kHz secular frequency. This is acceptable, however, it is on a similar scale as the observed heating rate observed in the AlN ceramic trap in Section 2. Placing the filter electronics on the trap chips or choosing a lower cut off frequency can reduce the Johnson noise.

\section{Integrating micro-optics in chip-based ion traps}
 
Integrated micro-optics for ion addressing and detection improve the scalability and reduce the system size. They can provide robust light delivery to multiple trap segments simultaneously or supply atom/light interfaces via optical cavities. In the \textit{IDEAL} project \cite{ideal}, we integrate micro-optics connected to optical fibers into a scalable chip ion trap. The trap geometry is reduced to a two chip 3D ion trap providing the same properties as the ion trap types presented in Section 2. For simultaneous readout of multiple trap segments, micro-optics will be used for optical detection and light delivery.  

Here, we study the effect of integrated micro-optics on the trapping fields. Optics for addressing the ions need  to be positioned in line of sight to the ions, however, surface charges on the micro-optics lead to additional electrical fields that can disturb the ions. To avoid dielectric surfaces, transparent conductive coatings, such as indium tin oxide (ITO), can be used to ground the surface \cite{Jansson2025}. Alternatively, grounded metallic mirrors can be used or the dielectric surface of an optic can be shielded with a conductive grounded housing to avoid distortions of the trapping fields. However, grounded components still have an effect on the electric fields of the trap which we examine with FEM simulations. 

Shifts in the radial direction can be compensated by applying an appropriate voltage to the DC and compensation electrodes. Residual RF fields in the axial direction lead to excess micromotion that cannot be compensated, and therefore need to be minimized. The residual micromotion limits spectroscopic measurements and can cause increased heating rates \cite{Wine1997, Kalin2021}. The ions are axially confined by a harmonic DC potential. The effect of the optics on the axial trapping potential was also examined in the simulations. 

\begin{figure}
	\centering
		\includegraphics[scale = 1]{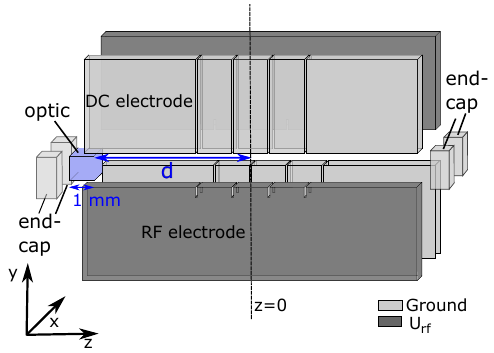}      
		\caption{\label{fig:fem}Setup for FEM simulations comprising four trap electrodes, two endcap electrodes, and micro-optics. Two of the electrodes are grounded and two are on a RF potential. The grounded electrodes have 0.04 mm wide slits that separate the 2 mm wide segments. The axial DC trapping potential is applied on the outer 2 mm segments. The blue cuboid represents integrated micro-optics. The origin of the coordinate system coincides with the ion´s position. }
\label{fig:FEM}
\end{figure}%
\begin{figure}
	\centering
		\includegraphics[scale = 0.35]{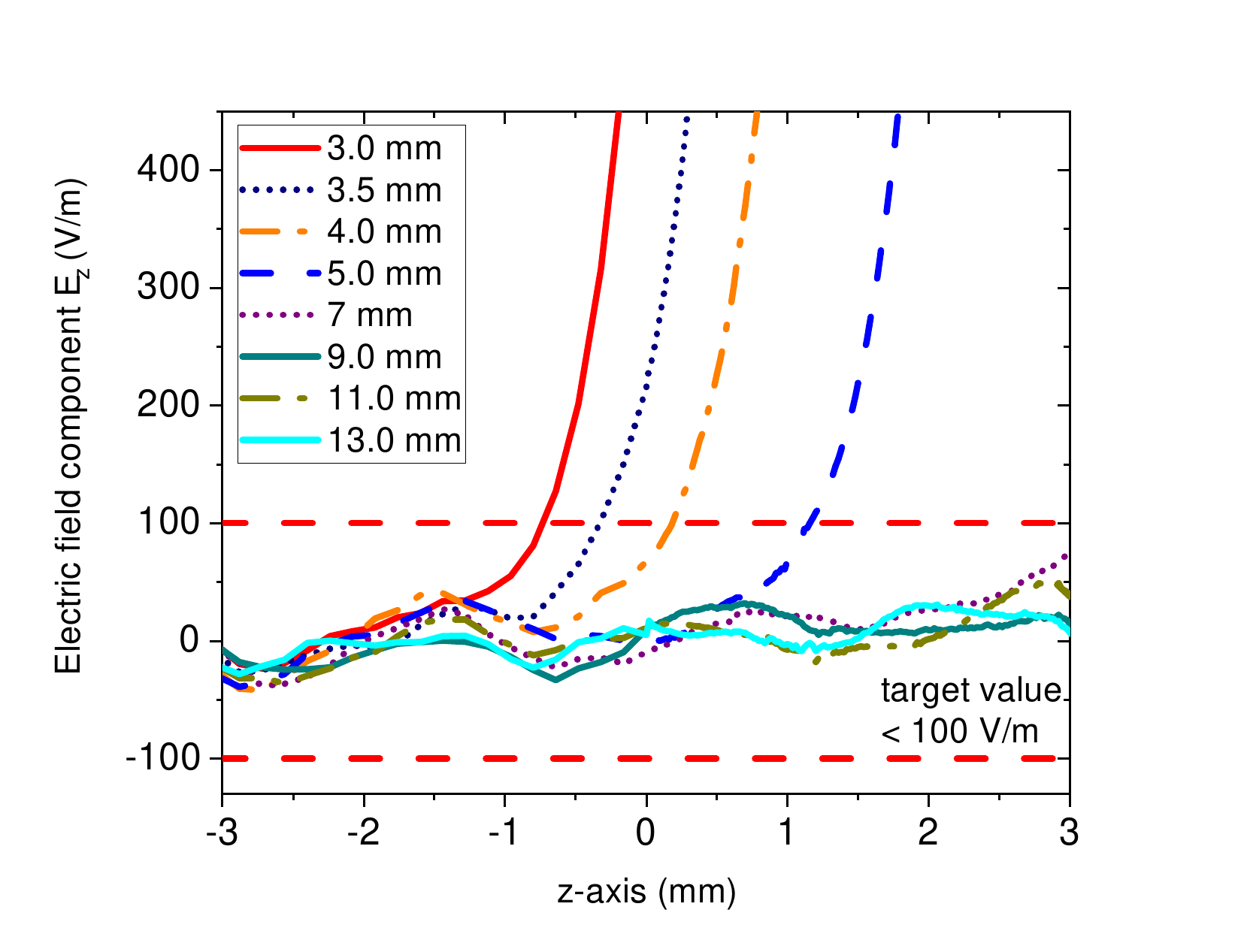}      
		\caption{\label{fig:axis} FEM calculation of the axial component of the RF field $E_z$ on the trap axis with an optic inserted into the trap on the trap axis. Different colors indicate different distances of the micro-optic from the position of the ion.}
\end{figure}

\subsection*{Simulation setup and parameters}

The FEM simulations are performed using a commercially available software (COMSOL Multiphysics 5.6). The simulation comprises 2 trap RF electrodes, 2 grounded trap electrodes, 2 endcap electrodes, and the respective micro-optics, as shown in Figure \ref{fig:fem}. The electrodes are modeled as cuboids. The dimensions of the RF electrodes are (length $\times$ height$\times$ width = 26 $\times$ 5 $\times$ 0.4 mm$^3$). The grounded electrodes are segmented into 5 segments by 0.04 mm wide slits. The 3 central segments are 2 mm wide. The cuboids representing the RF electrodes have 0.5 mm long and 0.04 mm wide slits that are aligned with the slits on the grounded electrodes. The endcap electrodes are modeled as pairs of cuboids with dimensions 1$\times$2$\times$0.4 mm$^3$. All the electrodes are surrounded by a grounded box of size 50$\times$50$\times$50 mm$^3$, so the box is at a distance of 10-30 mm from the trap electrodes. The tetrahedral mesh for the simulation is generated using the built-in routines of the software. To get a high resolution of the electric fields at the position of the ions, the mesh is chosen coarser at the electrodes and increasingly fine towards the trap center.  

In the simulations, the rf trap potential is considered at a fixed phase, which reduces the electric field calculation to a static problem. The potential on the rf electrodes is set to $U_{\mathrm{rf}}=1000$ V, the other electrodes are grounded. We chose a residual axial electric field amplitude of $E_z < 100$ V/m as the target value which is the same as in reference \cite{Hers2012}. 
In a separate simulation run, the axial confining potential was applied to the outer 2 mm segments and set to 1 V, while the RF electrodes and endcaps were grounded. The secular frequency for an ${}^{171}$Yb${^+}$ ion is calculated to evaluate the effect of the grounded optics. For the calculation, an RF frequency of $\Omega_{rf}= 2\pi\times 16$ MHz is assumed.

\subsection*{Micro-optics on the trap axis}

Micro-optics on the trap axis as shown in figure \ref{fig:fem}, for example for axial laser cooling, are approximated with a rectangular solid with 1.0 mm height, 1.0 mm length, and 0.8 mm width. The surface of the optic is grounded. This corresponds to one endcap moved in toward the center of the trap. The distance of the optic from the ions is varied between $d= 3$ mm and $d=13$ mm. The effect on the axial trapping fields of the grounded solid is found to be small, because the optic is outside the 2 mm segments. Even for the smallest distance to the ions of 3 mm the axial frequency of the harmonic secular motion is reduced by less than 0.5\%.

The effect on the RF field is more significant as shown in figure \ref{fig:axis}, in particular, when the dielectric is positioned between the RF electrodes. At a distance $d$ of 5 mm from the center of the trap, the axial component of the RF field has dropped to less than 100 V/m. For short distances there is also an increased gradient in the residual axial RF field at the position of the ion of $2.0\times 10^6$ V m$^{-2}$ at 3 mm distance compared to $3.2\times 10^4$ V m$^{-2}$ without optics. The increased gradient can lead to increased heating of the ion \cite{Kalin2021}.

\begin{figure}[ht]
\centering
		\includegraphics[scale =0.8]{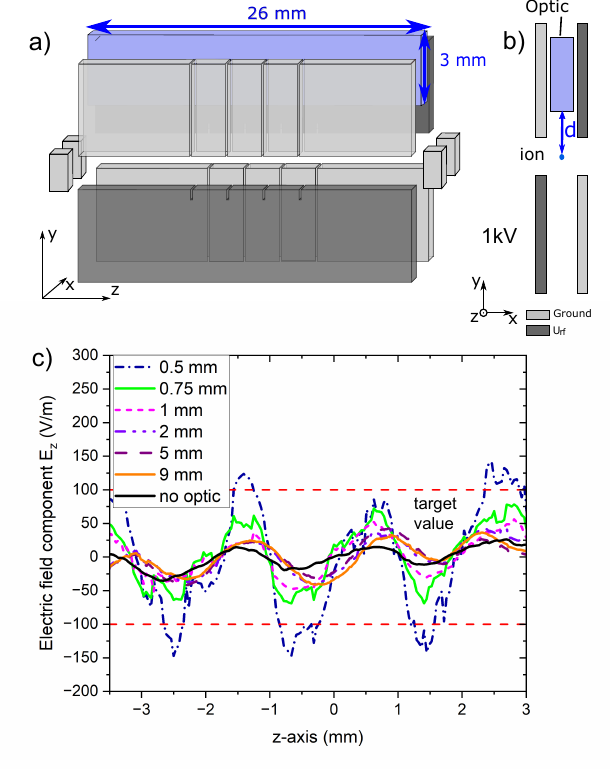}
		\caption{\label{FigYdirection}a) Schematic drawing of the trap with grounded optic inserted between the trap chips. b) Side view of the setup in a). c) FEM calculation of the axial component of the RF field $E_z$ on the trap axis.}
\end{figure}

\subsection*{Optics between the trap chips}

Another possibility is to insert optics between the trap chips from the y-direction perpendicular to the trap axis as indicated in figure \ref{FigYdirection}. For example, this can be optics for the laser addressing of ions in individual segments. The optics assembly is modeled as a cuboid as long as the rf electrode. In the simulation, the optics are placed at several distances between 0.5 mm and 9 mm from the ion's position. The results indicate that even at a distance of 0.5 mm, the axial component of the rf field (amplitude 1 kV) at the position of the ion is under 100 V/m (Figure \ref{FigYdirection} c). However, the gradient in the residual axial RF field of $2.7\times 10^5$ V m$^{-2}$ is increased compared to the trap without optical integration. The fluctuations in the simulation data on a sub-millimeter scale are due to the meshing of the FEM simulation. However, when positioned closer than 1.25 mm from the center of the trap, the grounded optics still shield the electric field of the electrodes which results in a distortion of the trapping fields. To quantify this shielding effect, we determine the axial trapping potential and observe a change of the secular frequency by more than 5 \%. For larger distances over 1.75 mm, this effect gets weaker and the change in frequency is under 1 \%.  

\begin{figure}[ht]
\centering
		\includegraphics[scale =0.9]{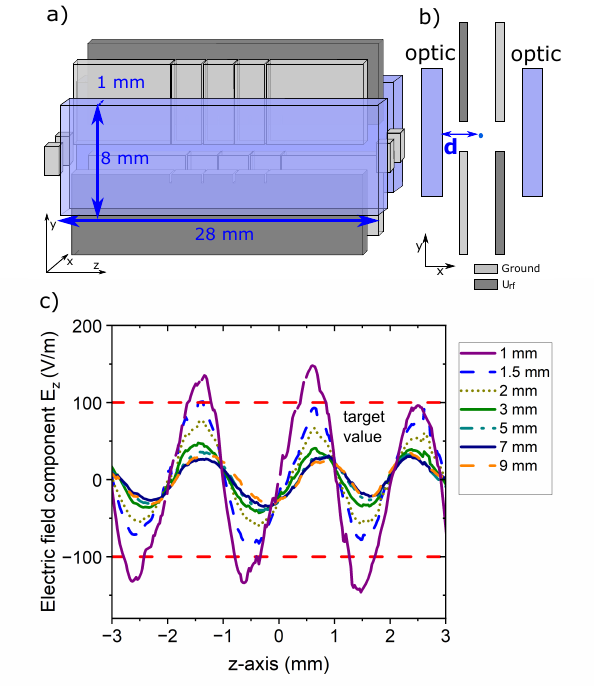}
		\caption{\label{FigDetection}a) Schematic drawing of the ion trap with grounded optics on both sides of the trap. This could be for example an objective for detection. b) Side view of the trap. c) FEM calculation of the axial component of the RF field $E_z$ on the trap axis.}
\end{figure}

\subsection*{Optics on both sides outside of the trap chip stack}

As a third possibility, a chip ion trap with optics on both sides outside of the trap chip stack is simulated as shown in figure \ref{FigDetection}. For example, this could be an objective lens and outcoupling optics for laser beams. The distance of the optics to the center of the trap is varied from $d=1$ mm to $d=9$ mm. From the simulation results, changes in the secular frequency of the ions and the electric field on the trap axis are determined. The secular frequency changes less than 1\% for all simulated distances, which is considered acceptable. The amplitude of the axial component of the RF field on the trap axis increases to over 100 V/m for distances $d <1.5$ mm, leading to field gradients of up to $2.8\times 10^5$ V m$^{-2}$ at the ion´s position. 

\begin{figure}
\centering
		\includegraphics[scale =0.32]{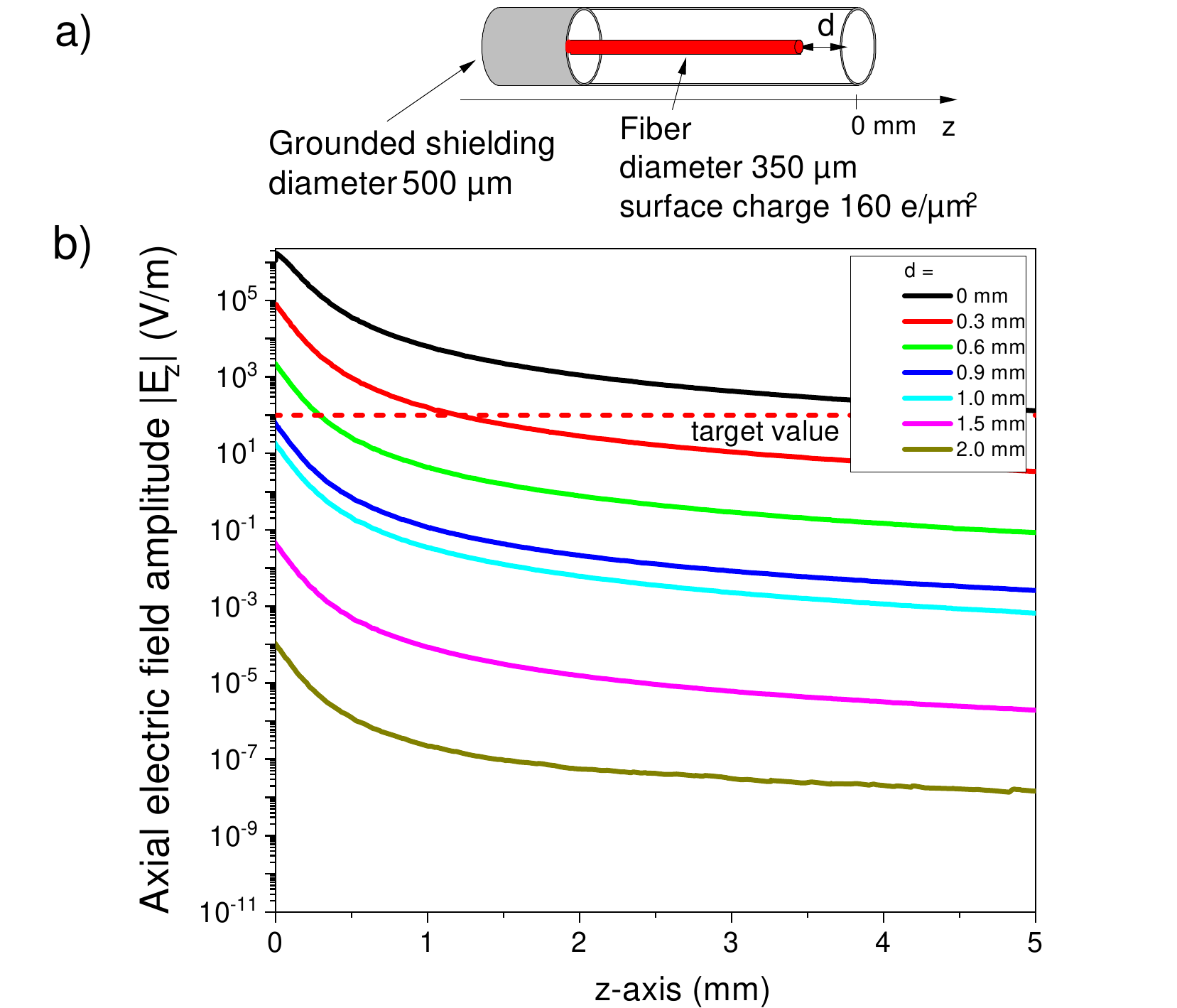}
		\caption{\label{Fig8} a) Schematic drawing of the fiber (red cylinder) in a grounded shielding (grey) which is partially transparent in this illustration to make the fiber visible. The shielding is protruding by the length $d$. b) FEM calculation of the axial electric field component $E_z$ on the z-axis as a function of the distance from the end of the shielding. Different lines belong to different lengths $d$ of the protrusion that are given in the legend. }
\end{figure}

\subsection*{Shielding}
Common ITO coatings may have low transmission for UV wavelengths.  Alternatively, an integrated fiber can be placed in a shielding, such as a protruding grounded metal cylinder (see figure \ref{Fig8}a). The simulations show how the electrical fields originating from the charged end of an optical fiber are shielded with a grounded cylinder as shown in figure \ref{Fig8}b. In the simulation, a fiber with 0.350 mm diameter and 5.00 mm length is placed in the center of a cylinder with an inner diameter of 0.5 mm and 10 mm length. The surface charge of the fiber is estimated to be 160 $e^-/ \mu$m$^2$ \cite{Ong2020}. The higher value compared to reference \cite{Ong2020} accounts for additional charging from UV light. The electrical field on the trap axis is evaluated for various positions of the fiber in the cylinder. The results in figure \ref{Fig8} suggest that with a protrusion of $d=1$ mm, the field outside the shielding is less than 100 V/m. In this case the electric field decreases to below 1 V/m at a distance of 0.5 mm from the cylinder. The simulation results indicate that it is possible to achieve efficient shielding of optical fibers without a transparent conductive coating.

\section{Summary}
Our 3D chip ion traps are constructed from two or four chips that contain the electrodes for ion trapping and for compensating parasitic electric fields. Chip ion traps made from PCB material are ideal for prototyping as they can be produced on a short time scale, offer low heating rates ($\dot{\bar{n}} = 1.2\pm 0.4$ s$^{-1}$ for $\nu_{\mathrm{sec}}=440$ kHz \cite{KellerProceedings2015}) and reach good vacuum pressure (e.g. $p<1\times10^{-10}$ mbar). Ion traps made from AlN ceramics or crystalline material or glass chips offer precise fabrication tolerances, which are valuable for applications that require low micromotion gradients and even lower heating rates ($\dot{\bar{n}} = 0.56\pm 0.06 $ s$^{-1}$ for $\nu_{\mathrm{sec}}=440$ kHz \cite{Kalin2021}). We estimate the upper limit for the heating rate of the ion caused by Johnson noise on a DC trap electrode at room temperature to be $\dot{\bar{n}} =0.31 $ phonon/s. Positioning RC filter electronics close to the electrode on the trap chip can help to reduce the Johnson noise on the trap electrodes. However for small chip traps, the filters may have to be positioned on the carrier board. In this case, an electric connection after the filter to the trap electrode with low resistance is advantageous. 

Using FEM simulations of a 3D chip ion trap with integrated micro-optics, we show that it is possible to insert micro-optics between the trap chips without significantly distorting the trapping fields given that the distances from the ions are chosen large enough. The micro-optic integration profits from the good electrical shielding of the 3D ion trap geometry. We quantify the distortion by setting limits for the axial component of the residual RF field and the change in secular frequency of a $^{171}$Yb$^+$ ion. Micro-optics can be introduced on the trap axis or from the sides of the trap without changing the secular frequencies by more than 1 \% at a distance $\geq 1.75$ mm. The necessary distance depends on the direction from which the optic is inserted. However, placing the optics between the RF electrodes causes an unwanted RF component in the axial direction with a field amplitude that exceeds the target value of $E_{\mathrm{rf,z}}=100$ V/m for distances $<5$ mm. Inserting the optics perpendicular to the trap axis causes considerably less distortion of the trapping fields, even at distances of 1 mm from the ion.
In conclusion, the integration of micro-optics improves the scalability of 3D ion traps and offers new possibilities for simultaneous manipulation and detection of multiple ion ensembles in the same trap. 

\subsection{Acknowledgments}
The authors thank C. Feist for the help with laser cutting, R. Mee{\ss}  for the fruitful discussion on trap fabrication, J. Keller for the discussion on experimental issues, R. Banerjee Chaudhuri for proofreading the manuscript,  M. Kazda for the support for the placement of SMD components, J. Blohm,  for the support with SEM. We acknowledge support from the German Federal Ministry of Education and Research for the project \textit{IDEAL} grant no. 13N14962 and the project \textit{opticlock} within the program “pilot projects quantum technologies” grant no. 3N14380. We acknowledge support by the Project 20FUN01 \textit{TSCAC}, which has received funding from the EMPIR programme co-financed by the participating states and from the European Union’s Horizon 2020 research and innovation programme, and by the Deutsche Forschungsgemeinschaft (DFG, German Research Foundation) under SFB 1227 DQ-mat—Project-ID No. 274200144—within Projects B02 and B03.  E. Jansson acknowledges support from the Max-Planck-RIKEN-PTB-Center for Time, Constants and Fundamental Symmetries.

\newpage
	\section*{References}
	\bibliographystyle{iopart-num}
	\bibliography{TrapPaperRef}

\end{document}